\documentclass[12pt]{article}

\title{}
\author{}
\date{}
\usepackage{amsmath}
\usepackage{mathtools} 
\usepackage{amssymb}
\usepackage{bm}
\usepackage{siunitx}
\usepackage{gensymb}
\usepackage[margin=1in]{geometry}
\usepackage[document]{ragged2e}
\usepackage{caption} 
\usepackage{graphicx}
\graphicspath{C:/Users/Daniel/Documents/Bornn_Lab_Summer_2018/arXiv_submission/Plots/}
\usepackage{grffile}
\usepackage{subfig}
\usepackage[T1]{fontenc}
\usepackage{mathptmx}
\usepackage{setspace}
\usepackage{hanging}
\title{Using In-Game Shot Trajectories to Better Understand Defensive Impact in the NBA}
\author{Daniel Daly-Grafstein$^1$ and Luke Bornn$^1$}
\date{%
    $^1$Simon Fraser University\\[2ex]%
    \today
}
\begin{document}
\maketitle

\textbf{Abstract} \, As 3-point shooting in the NBA continues to increase, the importance of perimeter defense has never been greater. Perimeter defenders are often evaluated by their ability to tightly contest shots, but how exactly does contesting a jump shot cause a decrease in expected shooting percentage, and can we use this insight to better assess perimeter defender ability? In this paper we analyze over 50,000 shot trajectories from the NBA to explain why, in terms of impact on shot trajectories, shooters tend to miss more when tightly contested. We present a variety of results derived from this shot trajectory data. Additionally, pairing trajectory data with features such as defender height, distance, and contest angle, we are able to evaluate not just perimeter defenders, but also shooters' resilience to defensive pressure. Utilizing shot trajectories and corresponding modeled shot-make probabilities, we are able to create perimeter defensive metrics that are more accurate and less variable than traditional metrics like opponent field goal percentage. \par

\section{Introduction} 

Perimeter defense in the NBA involves defenders attempting to stop, contest, or block outside jump shots by the opposing team. With three-point attempt rates continuing to rise, players' perimeter defensive ability is an important factor in determining a team's defensive success. However, it is difficult to quantify the ability of perimeter defenders. Additionally, while it is well-known that tightly contesting outside shots results in poorer shooting (Chang et al. 2014), little has been done to study why contesting shots decreases field-goal percentage (FG\%) and how contests affect the trajectory of shots. 

Defensive metrics are in general more difficult to measure and, traditionally, provide us less information than their offensive counterparts (Franks et al. 2015). Common box score metrics such as blocks and steals rely on discrete and easily countable events that do not provide us with a full picture of a player's defensive ability. Metrics like opponent FG\% and perimeter defense rating that try to quantify perimeter defense still rely on counting discrete events and can be highly variable. For example, players' opponent 3P\% (three-point percentage where the given player is the closest defender) has almost zero correlation year-to-year (Narsu 2017). Even commonly used advanced metrics like defensive rating and adjusted plus/minus do not give us information about why certain defenders are effective or not. With the introduction of player tracking data, a suite of new defensive metrics have been developed to try and fill the gap between offensive and defensive metrics (Franks et al. 2015, Goldsberry and Weiss 2013). While many of these new metrics do incorporate spatial player information, they still do not utilize the shot trajectory information given by the optical tracking data. Metrics that are based solely on binary make/miss shot information can be unstable, as a player's FG\% over a single season is inherently low sample size and may be highly variable (Daly-Grafstein and Bornn 2019). Additionally, these metrics still do not address the question of how contesting shots causes them to miss more frequently. 

In this paper we introduce a variety of results derived from shot trajectories in an attempt to quantify how contesting shots affects shooting percentage. We begin by using spatio-temporal information provided by optical tracking data to estimate shot trajectories and shot-make probabilities. We quantify each trajectory using three shot factor measures: depth, left-right distance, and entry angle (Daly-Grafstein and Bornn 2019, Marty 2018, Marty and Lucey 2017), and use these shot factors to model shot-make probabilities. Next, we pair defender and trajectory information to present a collection of results, including trajectory variation in relation to open vs. contested shots, and how defender height and distance affect shot angles and shot depths. In Section 3, we show using regression models that metrics derived from shot trajectory information stabilize inference, allowing us to estimate defender skill and shooter resiliency to defensive pressure in fewer games than previously possible.  \par

\section{Methods} 

\subsection{Dataset}

The data used for our analysis is the SportVu spatio-temporal tracking data provided by STATS LLC. This optical tracking data provides the $x$ and $y$ coordinates of the 10 players on the court and the $x$, $y$, and $z$ coordinates of the ball, 25 times per second. The data are also tagged with play-by-play event codes that indicate when events such as shots, dribbles, passes, etc. take place. We restrict our analysis to 50,916 three-point shots from the 2014-15 season. Following Daly-Grafstein and Bornn (2019), we now present a model for estimating shot-make probabilities. 

\subsection{Estimating Shot Trajectories}

To accurately estimate the ball's position near the basket, we fit a quadratic best fit line through the trajectory of each shot $i$ of the form: \par 
\vspace{-0.5cm}

\begin{equation}
E(Z_i) = \beta_0 + \beta_1x_i + \beta_2y_i + \beta_3x_i^2+ \beta_4y_i^2 +\beta_5x_iy_i 
\end{equation}

\vspace{-0.5cm}

We estimate the coefficients in (1) using a Bayesian regression with a conjugate Normal prior for $\beta$ of the form $\rho(\beta|\sigma^2,z,X) \sim N(u_{0}, \, \sigma^2\Lambda_{0}^{-1})$, and a conjugate inverse gamma prior $\rho(\sigma^2|z,X) \sim IG(a_{0},b_{0})$. The parameters of these priors are modeled using non-informative conjugate hyperpriors updated with pseudo-data reflecting our prior knowledge of shot trajectories. We specify 4 pseudo-data points: 2 set at the $x$, $y$ location of the shooter and 7 feet in height, and 2 set at the centre of the basket and 10 feet in height. After updates using the pseudo-data and optical tracking data, we take the posterior mean of $\beta$ as our estimate for the coefficients in (1) (Figure 1).

\begin{figure}[!t]
  \centering
  \includegraphics[width=10cm]{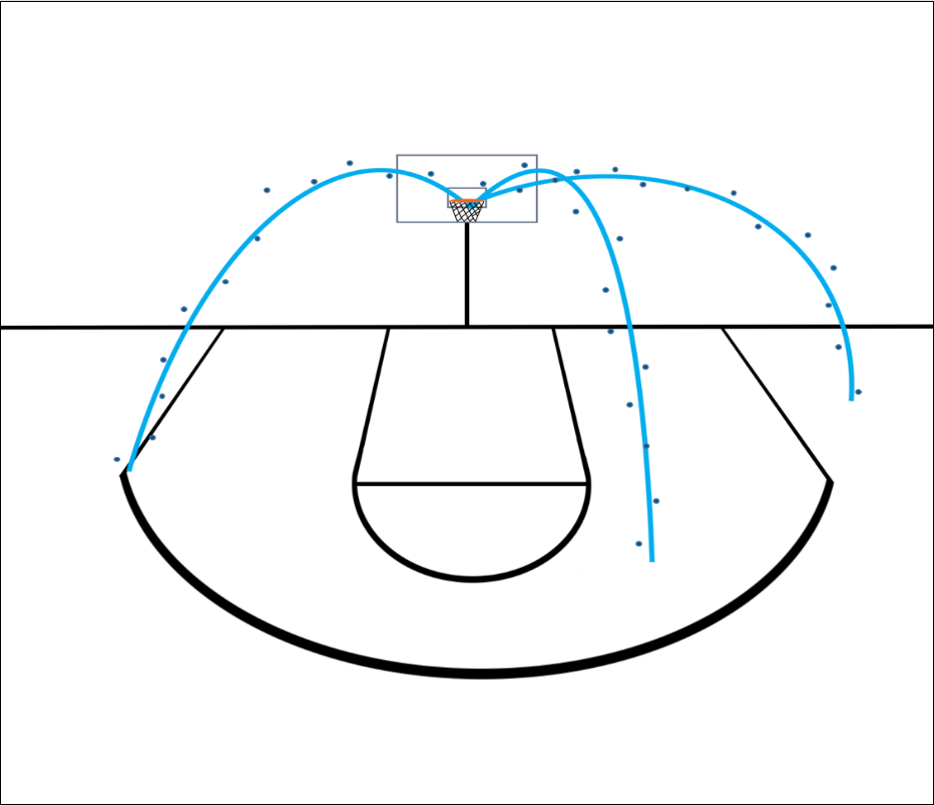}
  \caption{A graphical depiction of the shot trajectories from the SportVu database. The points represent data from the optical tracking database, while the smooth lines represent our modeled best-fit lines estimated using the Bayesian regression model (1). }
  \label{fig:trajectories}
\end{figure}
 
We then use (1) to calculate three shot factors for each trajectory - the shot depth, left-right distance, and entry angle - following the procedure of Marty and Lucey (2017). Shot depth is defined as the depth of the ball relative to the front rim as it enters the basket. Left-right distance is defined as the deviation of the ball from the center of the hoop. Entry angle is defined as the angle between the ball and the rim as it enters the basket. See Marty and Lucey (2017) and Daly-Grafstein and Bornn (2019) for further details.

\subsection{Modeling Shot-Make Probabilities}

The shot trajectories and derived shot factors described above give us more information on each shot than simply whether it is a make or a miss. If we summarize this trajectory information in a shot-make probability model, we can effectively Rao-Blackwellize shooting metrics and their derivatives by conditioning each shot's binary outcome on its make probability (Daly-Grafstein and Bornn, 2019). To accomplish this, we use the estimated shot factors described above as covariates in a logistic regression:
\vspace{-0.5cm}

\begin{equation} 
P(S_i=1)=\sigma \Bigg( \Bigg.
\begin{array}{c l}	
 &\beta_0 + \beta_1 \hat{D}_i + \beta_2 \hat{LR}_i + \beta_3 \hat{A}_i + \beta_4 \hat{D}_i^{2} + \beta_5 \hat{LR}_i^{2}\\
 &+\beta_6 \hat{A}_i^{2} +\beta_7 \hat{D}_i*\hat{LR}_i + \beta_8 \hat{D}_i*\hat{A}_i + \beta_9 \hat{LR}_i*\hat{A}_i 
\end{array}\Bigg. \Bigg)
\end{equation}
\vspace{-0.5cm}

with $P(S_i=1)$ representing the probability shot $i$ is a make, $\sigma(\text{x}) = \text{exp(x)}/(1+\text{exp(x)})$, and $\hat{D}_i$ , $\hat{LR}_i$, and  $\hat{A}_i$ representing the estimated depth, left-right distance, and entry angle of shot $i$, respectively. We train this model using 46,093 of the 50,916 threes from the 2014-15 NBA season, removing shot trajectories that were partially missing, especially noisy, or that resulted in modeled shot trajectories from (1) that were too far from the raw data. We show the distribution of modeled probabilities in relation to our three shot factors and the basket in Figure 2. \par

\begin{figure}[!tbp]
  \centering
  \subfloat[]{\includegraphics[width=6.8cm,height=5.4cm]{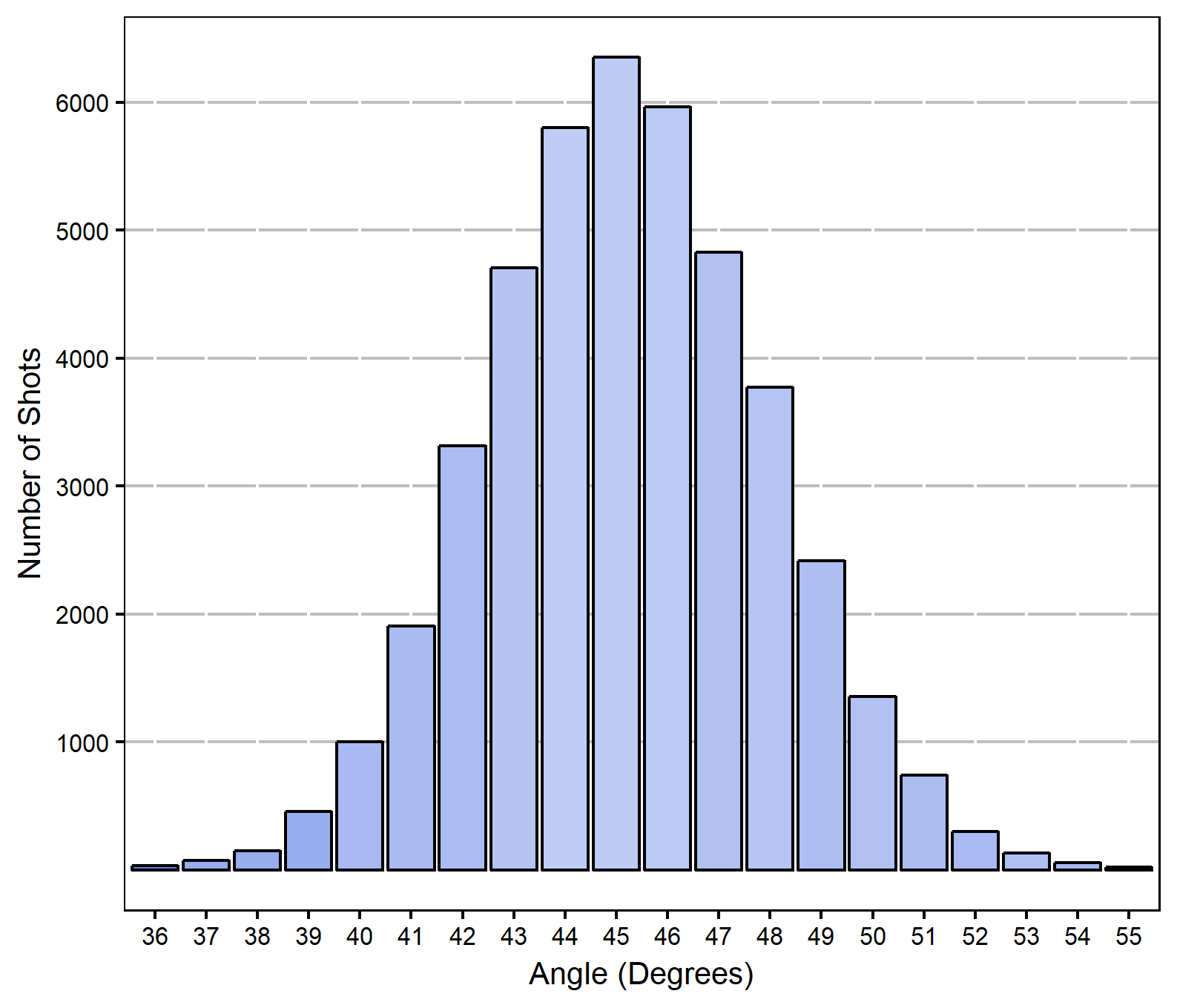}\label{fig:f3}}
  \subfloat[]{\includegraphics[width=7.5cm,height=6cm]{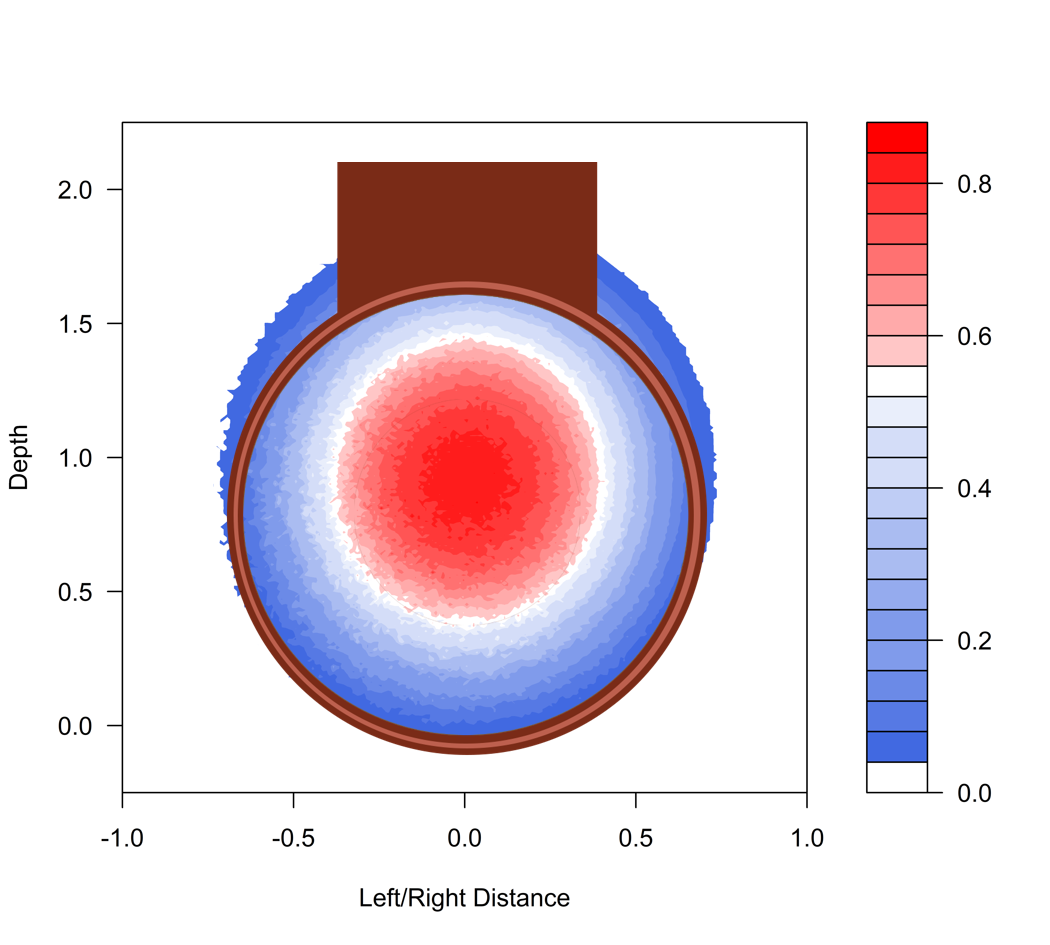}\label{fig:f3}}
  \caption{Figure (a) shows the mean predicted shot-make probabilities over the range of entry angles given by (2). Figure (b) shows the distribution of predicted shot-make probabilities over different values of shot depth and left-right accuracy in relation to the basket. Note the shot-make probability legend applies to both figures}
\end{figure}

\section{Results} 

\subsection{The Effect of Defenders on Shot Trajectories}

Here we present results based on shot trajectories that help give some insight into how exactly defending shots lowers shooting percentages. Firstly, when comparing open and contested 3-point shots, we find shots that are tightly contested have a 56\% larger variance in depth and a 38\% larger variance in left-right distance compared to open shots (Figure 3).  Contesting shots does not appear to introduce bias into the left-right accuracy of shooters, but does appear to cause shooters to bias their shots shorter than what is optimal. We also find that a smaller nearest defender distance (NDD) results in both higher entry angles and depths shorter in the hoop (Figure 4a, 4b). Additionally, conditional on defender distance, taller defenders result in higher shot trajectory angles when contesting 3-point shots (Figure 4a). The same trend is not as pronounced between defender heights and shot depths. Both our shot factors and those measured in Marty (2018) and Marty and Lucey (2017) using the Noah shooting system find that entry angles in the mid-40's result in the highest shooting percentage. Thus it appears that taller defenders are causing opponent shot trajectories to deviate from optimal angles. However, shooting percentages are more consistent over a range of entry angles compared to either left-right distance or shot depth, indicating the effect that taller defenders have on shot angles relative to overall shooting percentages may be minor. The more important effect may be how NDD affects shot depths. As in Marty and Lucey (2017), we find shot depths between 10" and 11" maximize 3P\%. In our dataset, shots landing at 9" depth are made at 60.1\% of the time, while shots landing at 10" depth are made 64.5\% of the time.  Thus, some of the drop in expected shooting percentage caused by contesting shots may be attributed to shooters biasing their shots shorter when confronted with tight defense. When looking at if defenders affected the left-right accuracy of shots, we do not find any effect of defender angle on shot trajectories. Specifically, defenders contesting from the left or the right of the shooter do not appear to bias shots in either direction.

\begin{figure}[!t]
  \centering
  \includegraphics[width=10cm]{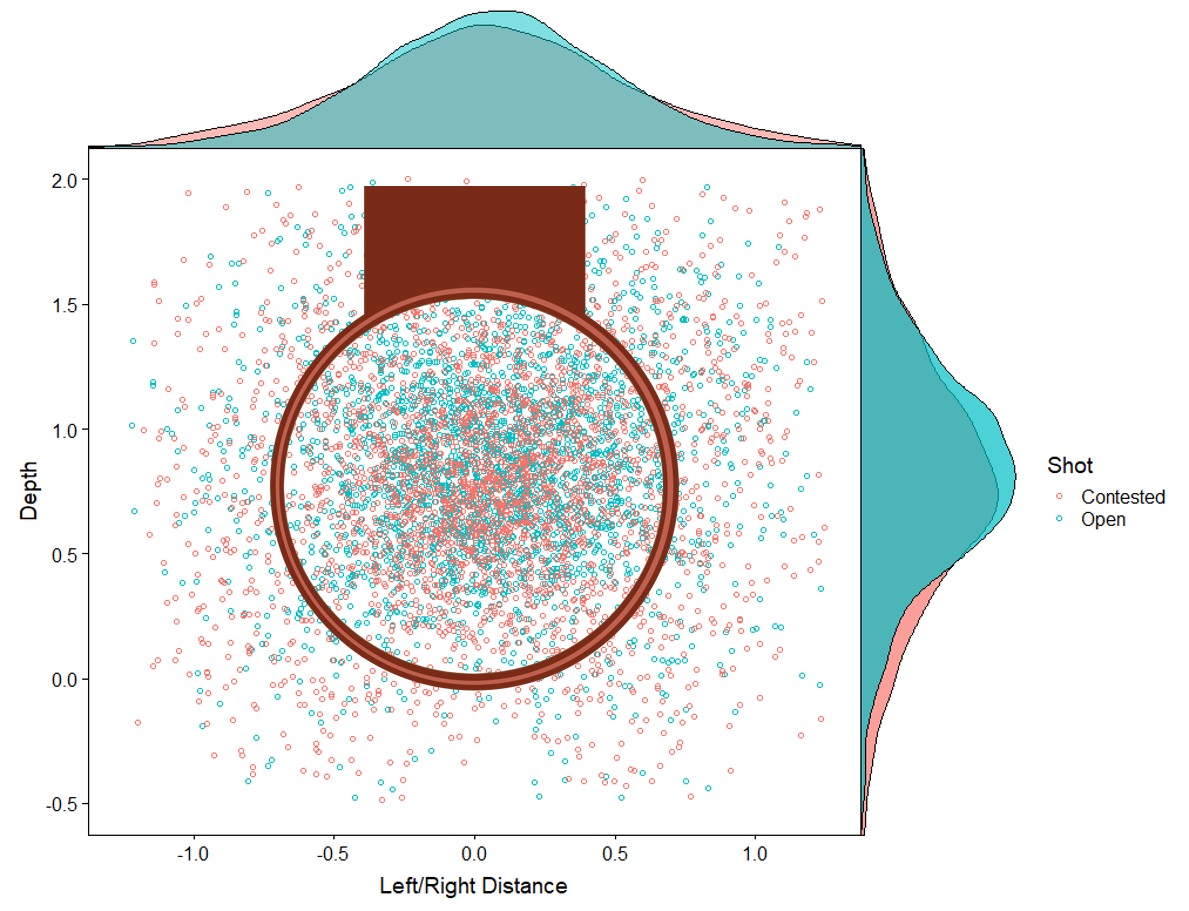}
  \caption{The distribution of open and contested 3-point attempts from the 2014-15 NBA season. Open and contested shots are defined as attempts with a NDD greater than 6 feet and less than 4 feet, respectively. Here NDD is taken as the distance of the closest defender to the shooter when the shot is released. Depth and left-right measurements are given in feet. }
  \label{fig:scattermarginal}
\end{figure}

\begin{figure}[!tbp]
  \centering
  \subfloat[]{\includegraphics[width=7.8cm,height=6.8cm]{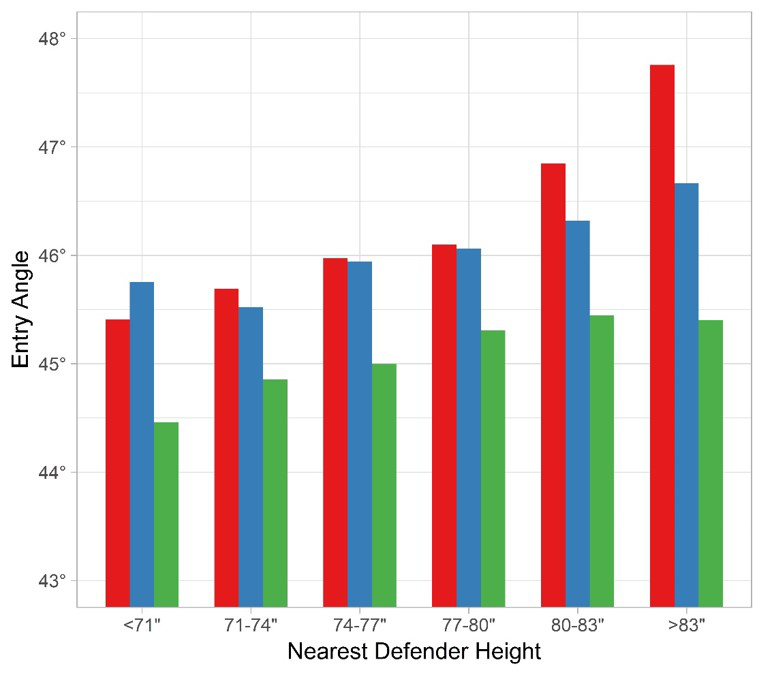}\label{fig:f4}}
  \subfloat[]{\includegraphics[width=8.3cm,height=6.8cm]{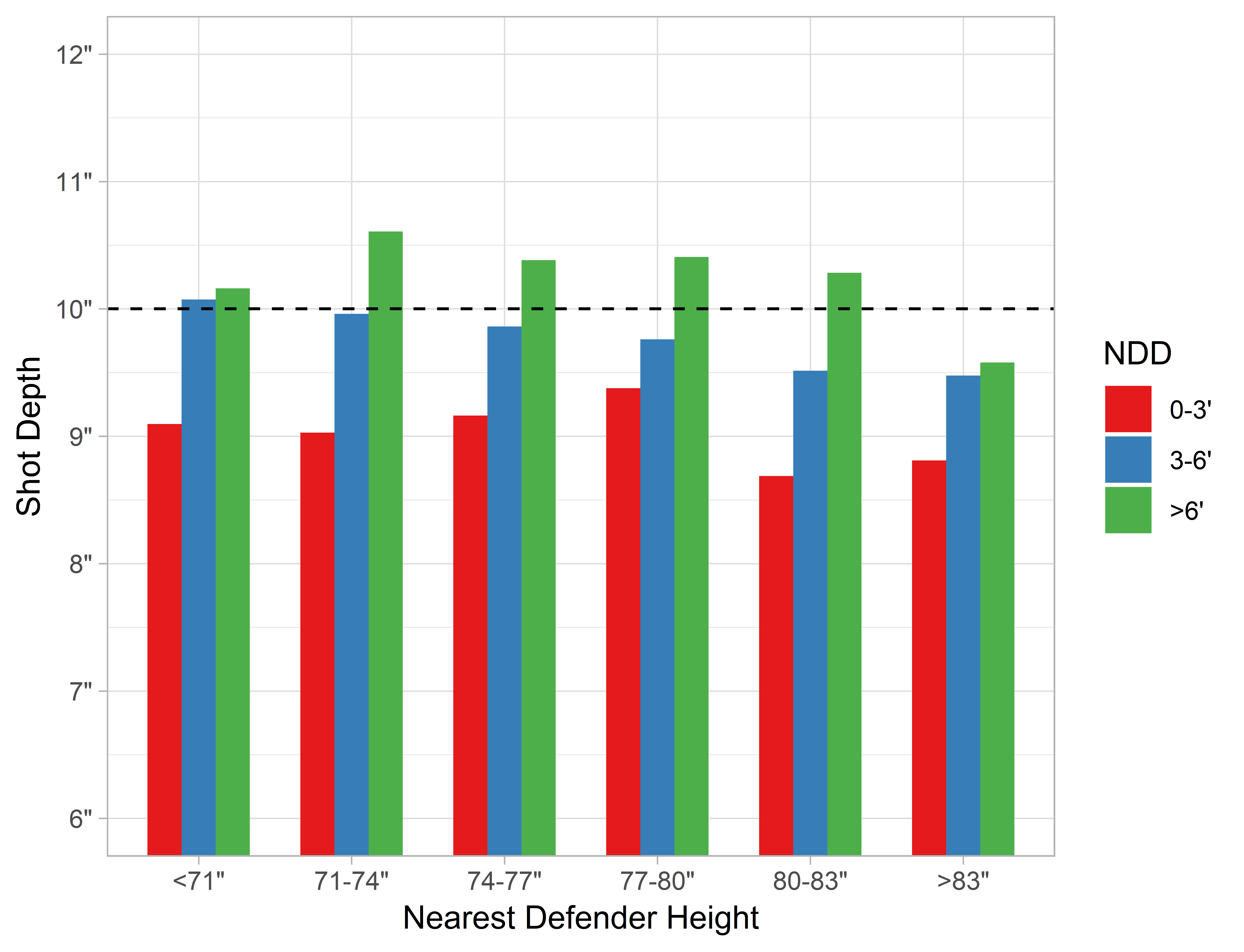}\label{fig:f4}}
  \caption{The entry angle (a) and shot depth (b) of all 3-point shot attempts during the 2014-15 season. Shot attempts are categorized by the nearest defender's distance (NDD) and the nearest defender's height. In Figure (b) the dotted horizontal line indicates the shot depth at which 3P\% is maximized. }
\end{figure}

\subsection{Evaluating Perimeter Defenders and Shooters}

As mentioned in Section 1, a player's opponent 3P\% is not a reliable perimeter defensive metric because it is quite variable, having almost no year-to-year correlation. Here we try to improve this metric by utilizing the modeled shot-make probabilities calculated in Section 2.2. To this end, we create 2 regression models to evaluate each player's defensive ability when they are tagged as the nearest defender. The first estimates the defensive impact of each player using make/miss indicators as the response (model 1), essentially giving the magnitude of difference between 3P\% when the defender of interest is defending compared to a weighted average of the offensive players' 3P\% over the season. The second model does similar, except uses shot-make probabilities as the response (model 2). These models have the form:  

\begin{equation}
Y_{ijk} = \beta_0 + \alpha_j + \gamma_k
\end{equation}

where $Y_{ijk}$ is the $i^{th}$ shot taken by the $j^{th}$ player and defended by the $k^{th}$ player.  $Y_{ijk}$ is either a binary indicator in the case of model 1, or the modeled shot-make probability of shot $i$ in the case of model 2. Using sum-to-zero contrasts, the $\alpha_j$'s are the differences between each player's 3P\% and the league average in the first model, and estimated differences between each player's mean shot-make probability and the league average shot-make probability in the second. Similarly, the $\gamma_k$'s are estimates of each defender's impact on opponent 3P\% in the first model, and estimates of each defender's impact on opponent three-point shot-make probability in the second model. 

If we consider the $\gamma_k$ values estimated using binary shot outcomes over the entire 2014-15 season as each player's true perimeter defensive impact, we can show that using shot-make probabilities allows us to estimate coefficients with less data than when using make/miss responses (Figure 5a). The MSEs of coefficients estimated using fewer than 50\% of the games from the 2014-15 season are smaller when using shot-make probabilities, and these gains are especially evident at low sample sizes. Additionally, we find that when comparing ranks of defenders from the first and second half of the 2014-15 season, coefficients estimated using shot-make probabilities outperform those estimated with make/miss outcomes in terms of consistency of player ranks ($\rho$ = 0.17 vs. 0.025, respectively). Thus, we can use our new metric to more accurately rank perimeter defenders compared to opponent 3P\% (Table 1).

 \begin{figure}[!t]
  \centering
  \subfloat[]{\includegraphics[width=7.8cm,height=6.8cm]{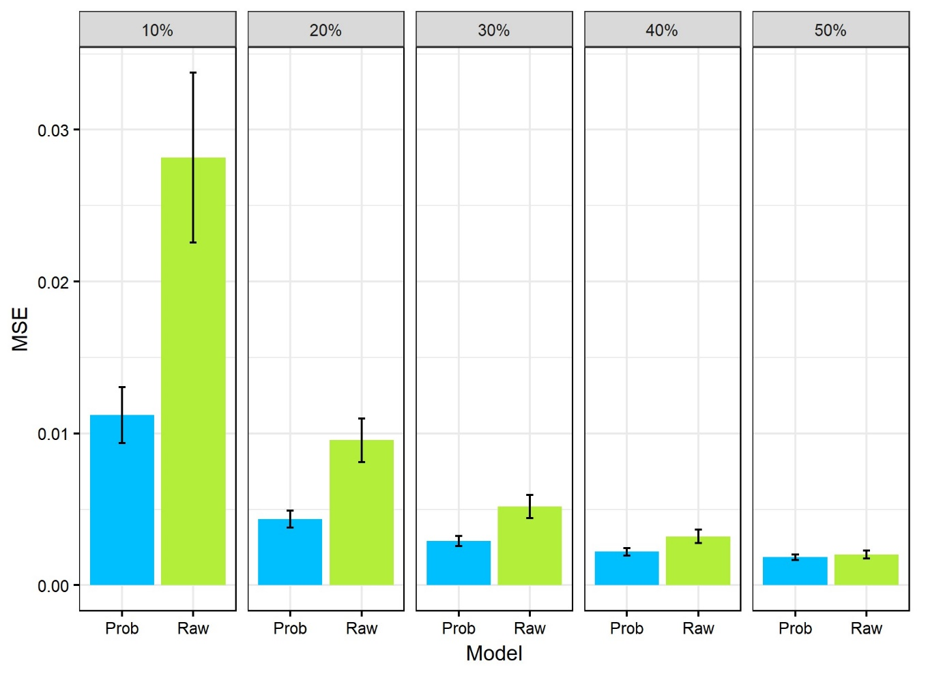}\label{fig:f5}}
  \subfloat[]{\includegraphics[width=8.3cm,height=6.8cm]{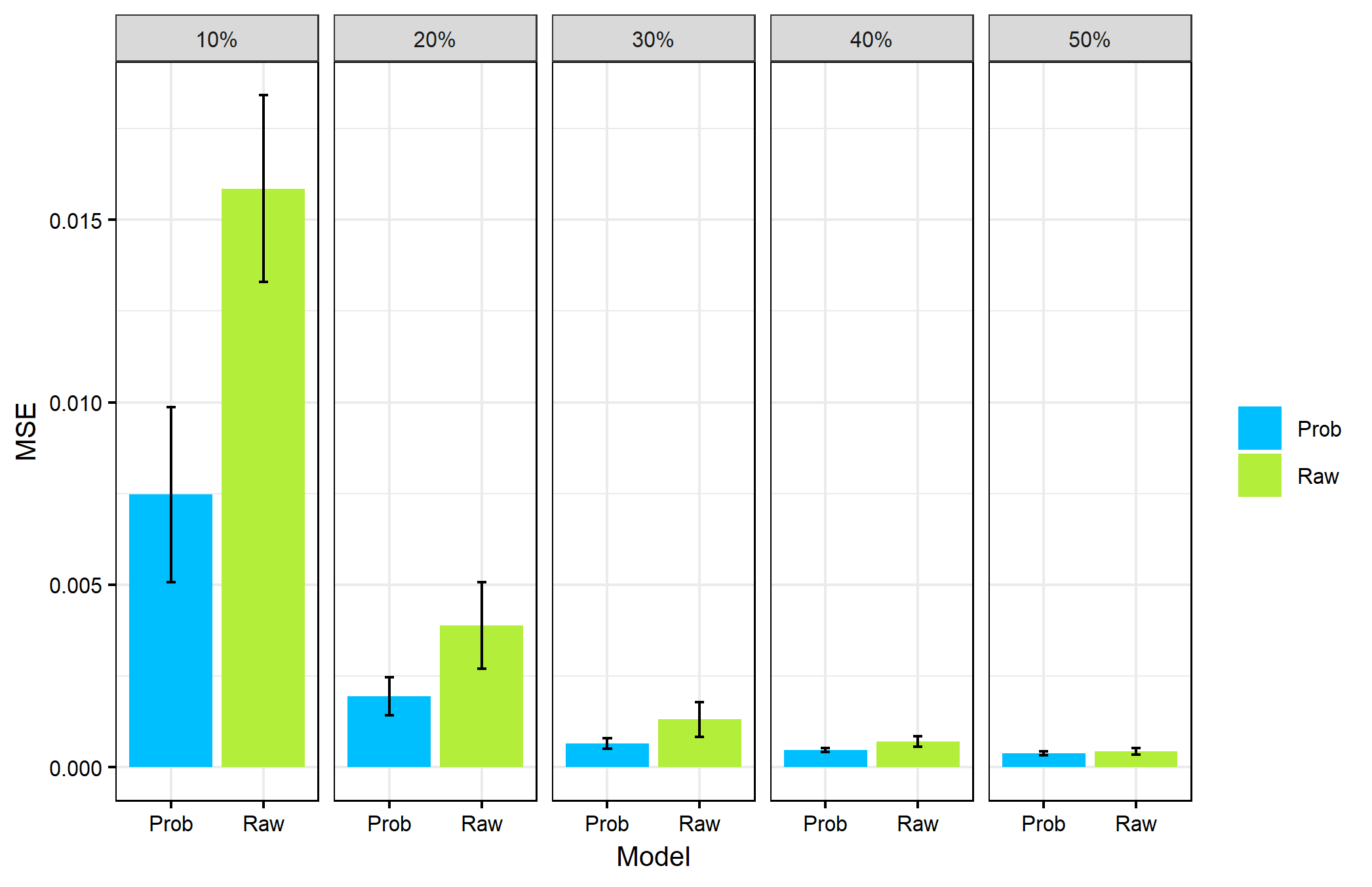}\label{fig:f5}}
  \caption{ Figure (a) depicts the mean squared error (MSE) of the $\gamma_k$'s from (3) estimated using 10\%, 20\%, 30\%, 40\%, and 50\% of the games in the 2014-15 season. Coefficients using model 1 (Raw) and model 2 (Prob) are compared to coefficients estimated using the entire 2014-15 season data and make/miss responses. These coefficients correspond to the defensive impact of each player. Figure (b) depicts the same MSE as (a) except the coefficients correspond to each shooter's interaction with NDD, denoted as $\gamma_j$ in (4).}
\end{figure}

%We can perform a similar analysis to measure how effective shooters are at responding to defensive pressure. We use a similar setup to the one summarized by (3), but this time the $\gamma_k$ coefficients denote interactions between the shooter and NDD, and in this case $j=k$ for all shots $i$. Again we have two models, the first using binary make/misses as the response, the second using estimated shot-make probabilities. These models now have the form: 

We can perform a similar analysis to measure how effective shooters are at responding to defensive pressure. We again create 2 regression models, this time to evaluate how players' shooting percentage changes based on nearest defender distance. The first model estimates the change in a player's 3P\% for every foot change in the NDD, while the second estimates the change in mean shot-make probability for every foot change in NDD. These models have the form:

\begin{equation}
Y_{ij} = \beta_0 + \alpha_j + \gamma_j*NDD_{ij}
\end{equation}

\newcommand\Tstrut{\rule{0pt}{2.6ex}}         % = `top' strut
\newcommand\Bstrut{\rule[-0.9ex]{0pt}{0pt}} 

\vspace{1cm}
\begin{table}[!b]
\caption{Nearest Defender Impact on Shots}
\centering
\begin{tabular}{|cccc|cccc|}
  \hline 
 Rank & Defender & $\gamma_k*100$ & Opp Prob &  Rank & Defender & $\gamma_k*100$  & Opp Prob\Tstrut\Bstrut\\
  \hline
  1 & Boris Diaw & -6.71 & 30.0\% & 137 & Derrick Williams & 8.57 & 45.8\%  \Tstrut\\ [1ex]
  2 & Draymond Green & -5.92 & 32.0\% & 136 & Channing Frye & 7.15 & 43.0\% \\ [1ex]
  3 & Langston Galloway & -5.25 & 30.6\% & 135 & Vince Carter &\ 5.96 & 41.7\% \\ [1ex]
  4 & Patrick Beverley & -4.55 & 31.9\% & 134 & Kirk Hinrich & 5.93 & 42.2\% \\ [1ex]
  5 & Wesley Johnson & -4.39 & 31.7\% & 133 & Jameer Nelson & 5.69 & 42.8\% \Bstrut\\ [0.5ex]
   \hline
\end{tabular}
\begin{flushleft}
The top and bottom perimeter defenders estimated via (3) using shot-make probabilities from (2). The $\gamma_k*100$ values represent the estimated difference in 3-point shot-make probability percentage  per 100 shots when the given player is the primary defender compared to a weighted average of probabilities based on their opponent's shooting skill. The Opp Prob column denotes the mean estimated shot-make probability of shots where player $k$ is the closest defender. Restricted to players who defended at least 100 three-point shots during 2014-15.  
\end{flushleft}
\end{table}
\vspace{0.1cm}

where $Y_{ij}$ is the $i^{th}$ shot taken by the $j^{th}$ player, and the $\alpha_j$'s are defined similarly to (3). The $\gamma_j$'s now denote the estimated interaction effect between each shooter and the NDD. Thus the $\gamma_j$ coefficients represent the estimated change in mean 3P\% (shot-make probability) for every one foot change in the NDD for each shooter. Again we find that we can estimate coefficients using less data (Figure 5b) and that shooter rankings are more consistent when using shot-make probabilities ($\rho$ = 0.20 vs. 0.033, respectively). Shooter rankings based on changes in shot-make probability are presented in Table 2. For example, Kemba Walker's estimated mean three-point shot-make probability decreases 1.98\% points less than the league average for every foot closer the nearest defender is.  

\vspace{1cm}
\begin{table}[!b]
\caption{Perimeter Shooter Resiliency to Shot Contests}
\centering
\begin{tabular}{|ccc|ccc|}
  \hline 
 Rank & Shooter & $\gamma_j*100$ &  Rank & Shooter & $\gamma_j*100$ \Tstrut\Bstrut\\
  \hline
  1 & Michel Carter-Williams & 3.45 & 137 & Aaron Brooks & -2.54 \Tstrut\\ [1ex]
  2 & Rasual Butler & 3.39 & 136 & Langston Galloway & -2.41 \\ [1ex]
  3 & Austin Rivers & 2.86 & 135 & Russell Westbrook &\ -2.37 \\ [1ex]
  4 & Kemba Walker & 1.98 & 134 & Nik Stauskas & -2.35 \\ [1ex]
  5 & Gerald Henderson & 1.58 & 133 & Rovert Covington & -2.02 \Bstrut\\ [0.5ex]
   \hline
\end{tabular}
\begin{flushleft}
The top and bottom shooters resilient to defensive pressure estimated via (4) using shot-make probabilities. Values represent the estimated change in each player's 3-point shot-make probability per 100 shots for every 1 foot decrease in NDD relative to the league average. Restricted to players who attempted at least 100 three-point shots during 2014-15. 
\end{flushleft}
\end{table}
\vspace{0.1cm}

\section{Discussion and Conclusion}

Substituting shot-make probabilities for binary make/miss outcomes is an example of Rao-Blackwellizing FG\%. If we model shots as Beta-Bernoulli random variables, shot-make probabilities become a sufficient statistic for shooting ability, and thus conditioning on these probabilities will, by the Rao-Blackwell theorem, result in an estimator with lower variance (Daly-Grafstein and Bornn 2019). The results presented in this paper are just a few examples of the improvements Rao-Blackwellization can give. With tracking data now available in hockey, football, and soccer, trajectory data can be leveraged to calculate similar goal/pass-make probabilities that may result in improvements similar to those seen in this paper. 

The results presented in Section 3 illustrate the improvements gained by using shot trajectories estimated from the tracking data to evaluate defender skill. We believe this work has opened up many areas of future research. For example, nearest defender distance is not the most reliable way to quantify the defensive pressure on a shot. It does not give us any indication of how the defender is oriented in relation to the shooter, and also may tag a player that is not the primary defender. We may be able to improve our defensive impact metric by using a more reliable measure of who the primary defender is (e.g. Franks et al. 2015), or by trying to incorporate the intensity of the defensive contest (e.g. Csapo and Raab 2014). Furthermore, we defined a relatively simple model in (3) that estimates a mean for each player's defensive impact. Conditioning on other covariates, such as shot location, shooter position, or even NDD, may give a more accurate estimation of players' perimeter defensive ability. Finally, opponent FG\%, and its counterpart based on shot-make probabilities defined in this paper, may themselves be flawed metrics in evaluating perimeter defense. These metrics do not take into account defenders who stopped opponents from attempting a shot, forced their opponent to pass or create a turnover, or even prevented the opponent shooter from receiving the ball altogether. Combining the metrics defined in this paper with those that account for how defenders affect shot volumes and efficiency over the course of an entire defensive possession (e.g. Franks et al. 2015) may give a fuller picture of a player's perimeter defensive ability. \par

In this paper we sought to provide new descriptions for how defenders affect shots as well as construct metrics that are better able to estimate perimeter defender and shooter behavior. Following Marty and Lucey (2017), we presented a variety of results derived from shot trajectories. Similar to Marty and Lucey (2017), we found that three-point probabilities are highest at a depth of 10", and shots have a fairly consistent make probability over a range of entry angles. Additionally, we found that NDD increases variability in shot depth, while also biasing shots short. However, neither NDD nor defender angle seemed to bias the left-right location of shot trajectories, with NDD only increasing its variability. Thus it appears players are shooting with sub-optimal shot depths when facing defensive pressure. This may give players that train to correct this bias an opportunity to improve their three-point shooting.  Furthermore, our new metrics based on make-probabilities decreased the variation in estimation relative to their raw counterparts. These metrics may allow coaches to more accurately assess a player's perimeter defense, as well as indicate which outside shooters are most affected by tight defensive pressure. Teams could use this information to make better decisions about which players to guard on the three-point line, or to better evaluate their players' shot selection based on defensive pressure.    \par

\section*{References}
\singlespacing
\setlength{\parskip}{1em}
\begin{hangparas}{.5in}{1}

Chang, Y.H., R. Maheswaran, J. Su, S. Kwok, T. Levy, A. Wexler, and K. Squire. 2014. "Quantifying Shot Quality in the NBA." \textit{Proceedings of the 2014 MIT Sloan Sports Analytics Conference}.

Csapo, P. and M. Raab. 2014. "Hand down, man down. Analysis of defensive adjustments in response to the hot hand in basketball using novel defense metrics." \textit{PLoS ONE} 9(12): Retrieved 19 Feb. 2019, from https://doi.org/10.1371/journal.pone.0114184. 

Daly-Grafstein, D. and L. Bornn. 2019. "Rao-Blackwellizing field goal percentage." \textit{Journal of Quantitative Analysis in Sports} 0(0): Retrieved 19 Feb. 2019, from doi:10.1515/jqas-2018-0064.

Franks, A., A. Miller, L. Bornn, and K. Goldsberry. 2015. "Characterizing the spatial structure of defensive skill in professional basketball." \textit{Annals of Applied Statstics} 9(1): 94-121.

Franks, A., A. Miller, L. Bornn, and K. Goldsberry. 2015. "Counterpoints: Advanced defensive metrics for NBA basketball." \textit{Proceedings of the 2015 MIT Sloan Sports Analytics Conference}.

Goldsberry, K., and E. Weiss. 2013. "The Dwight effect: A new ensemble of interior defense analytics for the NBA." \textit{Proceedings of the 2013 MIT Sloan Sports Analytics Conference}.

Marty, R. 2018. "High-resolution shot capture reveals systematic biases and an improved method for shooter evaluation." \textit{Proceedings of the 2018 MIT Sloan Sports Analytics Conference}.

Marty, R. and S. Lucey. 2017. "A data-driven method for understanding and increasing 3-point shooting percentage." \textit{Proceedings of the 2017 MIT Sloan Sports Analytics Conference}.

Narsu, K. 2017. "Shot defense and separating metrics from actions." htpps://fansided.com/2017/01/12/nylon-calculus-shot-defense-metrics-actions/. Accessed December 3rd, 2018.

\end{hangparas}
\end{document}